\documentclass[prd,
twocolumn,
showpacs,preprintnumbers,nofootinbib]
{revtex4}
\usepackage{hyperref}
\usepackage{graphicx}
\usepackage{dcolumn}
\usepackage{bm} 
\usepackage{amsmath, amssymb, amsfonts}
\usepackage{multirow}
\usepackage{array}
\usepackage{longtable,lscape}

\newcommand{\GEV}{\mbox{GeV}}
\newcommand{\vecc}[1]{\mbox{\boldmath $#1$}}
\newcommand\ba{\begin{eqnarray}}
\newcommand\ea{\end{eqnarray}}
\def\be{\begin{equation}}
\def\ee{\end{equation}}
\newcommand{\bc}{\begin{center}}
\newcommand{\ec}{\end{center}}

\begin{document}

\title{On the Transfer of Polarization from the Initial
to the Final Proton \\ in the Elastic Process $e \vec p \to e \vec p$}

\author{M.~V.~Galynskii\footnote{galynski@sosny.bas-net.by}}

\affiliation{Joint Institute for Power and Nuclear Research -- Sosny, National Academy of Sciences of Belarus,
 220109 Minsk, Belarus}

\begin{abstract}
The $Q^2$ dependence of the ratio of the cross sections with and without proton
spin flip, as well as the polarization asymmetry in the process $e \vec p \to e \vec p$
has been numerically analyzed using the results of JLab's polarization experiments on the
measurements of the ratio of the Sachs form factors in the $\vec e p \to e \vec p$ process.
The calculations have been made for the case where the initial (at rest) and final protons
are fully polarized and have a common spin quantization axis, which coincides with
the direction of motion of the final proton. The longitudinal polarization transfer
to the proton has been calculated in the case of the partially polarized initial proton
for a kinematics used in the experiment reported in [A. Liyanage et al. (SANE Collaboration),
Phys. Rev. C {\bf 101}, 035206 (2020)], where the double spin asymmetry was measured
in the $\vec e \vec p \to e p$ process. A noticeable sensitivity of the polarization
transfer to the proton to the form of the $Q^2$ dependence of the ratio $\mu_p G_E/G_M$
has been found. This sensitivity may be used to conduct a new independent experiment to
measure this dependence in the $ e \vec p \to e \vec p$ process. A criterion to assess
the reliability of measurements of the ratio of Sachs form factors using the Rosenbluth
technique has been proposed and used to analyze the results of two experiments.
\end{abstract}

\maketitle

{\bf Introduction.}
The electric $G_{E} $ and magnetic $G_{M}$ form factors, the so-called Sachs
form factors (SFFs), have been experimentally studied since the mid-1950s
in the elastic scattering of unpolarized electrons off a proton. All experimental
data on the behavior of the SFFs have been obtained by applying the Rosenbluth technique
(RT), which is based on using the Rosenbluth cross section (in the one-photon
approximation) for the $ep \to ep$ process  in the rest frame of the initial proton \cite{Rosen}:
\ba
\label{Ros}
\sigma=\frac{d\sigma} {d\Omega_e}= \frac{\alpha^2E_2\cos^2(\theta_e/2)}{4E_1^{3}\sin^4(\theta_e/2)}
\frac{1}{1+\tau_p} \left(G_E^{2} +\frac{\tau_p}{\varepsilon}G_M ^{2}\right).
\ea
Here, $\tau_p=Q^2/4M^2,  Q^2=-q^2=4E_1 E_2\sin^2(\theta_e/2)$ is the square
of the momentum transfer to the proton; $M$ is the mass of the proton;
$E_1$ and $E_2$, are the energies of the initial and final electrons,
respectively; $\theta_e$ is the electron scattering angle;
$\varepsilon=[1+2(1+\tau_p)\tan^2(\theta_e/2)]^{-1}$ is the
degree of linear polarization of the virtual photon \cite{Dombey,Rekalo74,AR,GL97}
that varies in the range $0 \leqslant \varepsilon \leqslant 1$; and $\alpha=1/137$ is
the fine structure constant.

Formula (\ref{Ros}) shows that the main contribution to the cross section
for the process $ep \to ep$ at high $Q^2$ values comes from the term proportional
to $G_M^{\,2}$. Because of this, extracting the contribution of $G_E^{\,2}$  even at
$Q^2\geqslant 1$ GeV$^2$ becomes a challenging problem \cite{ETG15,Punjabi2015}.
The RT was used to determine the experimental dependence of SFFs on $Q^2$,
which is described up to $Q^2\approx6$ GeV$^2$ by the dipole approximation;
their ratio
\be
R \equiv \mu_p G_E/G_M
\label{Rffs}
\ee
is quite accurately approximated by the equality $R \approx 1$,
where $\mu_p$=2.79 is the magnetic moment of the proton.

Akhiezer and Rekalo \cite{Rekalo74} proposed a method to measure
the ratio $R$ that is based on the polarization transfer from the initial
electron to the final proton in the process $\vec e p \to e \vec p$. Precision
experiments based on this method conducted at JLab \cite{Jones00,Gay01,Gay02}
revealed that the ratio $R$ decreases rapidly with increasing $Q^2$, which indicates
that the dipole dependence (scaling) of the SFFs is violated. This
decrease proved to be linear in the range $0.4 \, \GEV^2 \leqslant Q^2 \, \leqslant 5.6$ \GEV$^2$.

Measurements of $R$, which were repeated with a higher accuracy
\cite{Pun05,Puckett10,Puckett12,Puckett17,Qattan2005} in a broad $Q^2$ range up to
8.5 GeV$^2$ using both the Akhiezer -- Rekalo method \cite{Rekalo74}
and the RT, only confirmed these disagreements. 

Experimental values of $R$ \cite{Liyanage2020} were obtained by the SANE
collaboration using the third approach. Namely, they were extracted from the measurements
of the double spin asymmetry in the process $\vec e \vec p \to e p$
in the case where the electron beam and the proton target are partially polarized.
The degree of polarization of the proton target was $P_t=(70 \pm 5)$\%.
The experiment was carried out at two electron beam energies
$E_1= 5.895$ and 4.725 GeV and two values of $Q^2 =2.06$ and 5.66 GeV$^2$.
The values of $R$ obtained in \cite{Liyanage2020} are
in good agreement with the results of previous polarization
experiments carried out in 
\cite{Jones00,Gay01,Gay02,Pun05,Puckett10,Puckett12,Puckett17}.

The fourth method was proposed in \cite{JETPL18}. This method allows extracting
$G_E^{\,2}$ and $G_M^{\,2}$ from the direct measurements of cross sections
with and without proton spin flip in the elastic process with the polarization transfer
from the initial to the final proton:
\ba
e(p_1)+ \vec{p}\,(q_1,s_1) \to e(p_2)+\vec{p}\,(q_2,s_2),
\label{EPEP}
\ea
when the initial (at rest) proton is fully polarized along the direction
of motion of the detected final recoil proton. This method, which is also applicable in the
two-photon exchange (TPE) approximation, enables measurement of the squares of absolute values
of the generalized SFFs in a similar way \cite{JETPL19}.

In this study, we use the results of JLab’s polarization experiments,
where the ratio $R$ was measured in the process $\vec e  p \to e \vec p$,
to numerically analyze the $Q^2$ dependence of the ratio of the cross sections with and
without proton spin flip and the polarization asymmetry in the process
$e \vec p \to e \vec p$ in the case where the initial (at rest) and final
protons are fully polarized and have a common spin quantization axis that coincides with
the direction of motion of the detected final recoil proton. In the case
of the partially polarized initial proton, the longitudinal polarization transfer to the
proton was calculated in the kinematics of the experiment
\cite{Liyanage2020}. A criterion to assess the reliability of the
measured $R$ ratio using the RT is proposed, which is used to analyze
the measurements made in two well-known experiments \cite{Andivahis1994,Qattan2005}.

{\bf Cross section for the process $e \vec p \to e \vec p$ in the rest frame
of the initial proton.} Let us consider spin 4-vectors $s_{1}$ and $s_{2}$
of the initial and final protons with 4-momenta $q_{1}$ and $q_{2}$, respectively,
in process (\ref{EPEP}) in an arbitrary reference frame. The conditions of orthogonality
($s_{i} q_{i} = 0$) and normalization ($s_{i} ^{2} = - 1$) unambiguously provide the following
expressions for the time ($s_{i0}$) and space ($\vecc s_i$) components of these spin
4-vectors $s_i=(s_{i0}, \vecc s_i)$ in terms of their 4-velocities $v_i=q_i/M$ ($i=1, 2$):
\ba
s_i=(s_{i0}, \vecc s_i), \; s_{i0}=\vecc v_i\, \vecc c_i, \;
\vecc s_i =\vecc c_i + \frac{(\vecc c_i \vecc v_i)\,\vecc v_i}{1+v_{i0}}\;,
\label{spinv}
\ea
where the unit 3-vectors $\vecc c_i$ ($\vecc c_i^{2}=1$) specify the spin
projection (quantization) axes.

In the laboratory reference frame, where $q_1=(M,\vecc 0)$
and $q_2=(q_{20}, \vecc q_2)$, the spin projection axes $\vecc c_{1}$
and $\vecc c_{2}$ are chosen such that they coincide with the
direction of motion of the final proton:
\ba
\vecc c = \vecc c_{1} =\vecc c_{2}=\vecc n_2=  \vecc {q_2}/|\vecc q_2|\,.
\label{LSO}
\ea
Then the spin 4-vectors $s_{1}$ and $s_{2}$ of the initial and final protons,
respectively, in the laboratory reference frame have the form
\ba
\label{DSB_LSO1}
s_1=(0,\vecc n_2 )\,, \; s_2= (|\vecc v_2|, v_{20}\, \vecc {n_2})\,,
\,\vecc n_2=  \vecc {q_2}/|\vecc q_2|\,.
\ea
Method \cite{JETPL18} is based on the formula for differential
cross section for process (\ref{EPEP}) in the laboratory reference
frame in the case where the initial and final protons
are polarized and have a common axis of the spin projections
$\vecc c$ given by Eq. (\ref{LSO}):
\ba
\label{RosPol}
\frac{d\sigma_{\delta_1, \delta_2}} {d\Omega_e}&=&
\omega_{+} \sigma^{\uparrow\uparrow}+\omega_{-}\sigma^{\downarrow\uparrow}\,,\\
\label{RosPol2}
\sigma^{\uparrow\uparrow}&=&\sigma_M \, G^2_E ,\;\;
\sigma^{\downarrow\uparrow}=\sigma_M \frac{\tau_p}{\varepsilon} \, G^2_M\,,\\
\sigma_M&=& \frac{\alpha^2E_2\cos^2(\theta_e/2)}
{4E_1^{\,3}\sin^4(\theta_e/2)} \frac{1}{1+\tau_p}\,. 
\ea
Here, $\omega_{\pm}$ are the polarization factors specified by the
formulas
\ba
\omega_{+}=(1 + \delta_1 \delta_2)/2, \, \,\omega_{-}=(1 -\delta_1 \delta_2)/2\,,
\label{omegi}
\ea
where $\delta_{1,2}$ are the double projections of the spins of the
initial and final protons on the common axis of spin
projections $\vecc c$ (\ref{LSO}). It should be noted that Eq. (\ref{RosPol}) is
valid at $-1\leqslant \delta_{1,2}\leqslant 1$.

The corresponding experiment to measure the squares of the SFFs in the processes with and without
proton spin flip can be implemented in the following way. The initial proton at rest
should be fully polarized along the direction of motion of the detected final
recoil proton. By measuring the differential cross sections $\sigma^{\uparrow\uparrow}$
and $\sigma^{\downarrow\uparrow}$ (\ref{RosPol2}) as functions of $Q^2$, it is possible to
determine the $Q^2$ dependence of $G_E^2$ and $G^2_M$ and to
measure in this way these form factors.

It is noteworthy that Eq. (\ref{RosPol}), similar to Eq. (\ref{Ros}), is the sum
of two terms one of which contains only $G_E^{\,2}$ and the second contains
only $G_M^{\,2}$. By averaging and summing Eq. (\ref{RosPol}) over the polarizations
of the initial and final protons, the Rosenbluth cross section given
by Eq. (\ref{Ros}) and denoted as $\sigma_R$ 
can be represented in the alternative form \cite{JETPL18}
\ba
\label{Ross}
\sigma_R =\sigma^{\uparrow\uparrow} + \sigma^{\downarrow\uparrow}\,.
\ea
Consequently, the two terms one of which contains only $G_E^{~2}$ and the second
contains only $G_M^{~2}$ in the representation of Rosenbluth formula (\ref{Ros})
in the form of Eq. (\ref{Ross}) physically mean the cross sections with and
without proton spin flip, respectively, in the case
where the initial proton at rest is fully polarized along
the direction of motion of the final proton.

It is often asserted in publications, including textbooks on elementary particle
physics, that the SFFs are employed simply because of convenience since
they enable the representation of the Rosenbluth formula in a straightforward
and concise form. Since such formal conclusions regarding advantages of using
the form factors are also contained in popular monographs published many
years ago \cite{AB,BLP}, they are not subject to criticism and are still
reproduced in the literature; see, e.g., \cite{Paket2015}.

Cross section (\ref{RosPol}) can be represented in the form
\ba
&& d\sigma_{\delta_1, \delta_2}/ d\Omega_e
\label{sigma_d1_d2}
=(1+\delta_2 \delta_f) (\sigma^{\uparrow\uparrow}+\sigma^{\downarrow\uparrow}),\\
\label{delta_f}
&& \delta_f=\delta_1 (R_{\sigma}-1)/(R_{\sigma}+1), \\
&& R_{\sigma}=\sigma^{\uparrow\uparrow}/\sigma^{\downarrow\uparrow},
\label{R_sig}
\ea
where $\delta_f$ is the degree of the longitudinal polarization
of the final proton. If the initial proton is fully polarized
($\delta_1=1$), $\delta_f$ coincides with the standard definition
of polarization asymmetry
\ba
\label{Asim}
A= (R_{\sigma}-1)/(R_{\sigma}+1)\;.
\ea

As follows from Eq. (\ref{RosPol2}), the ratio of the cross sections
with and without proton spin flip $R_{\sigma}$ (\ref{R_sig}) can be
represented in terms of the experimentally measurable
quantity $R \equiv \mu_p\, G_E/G_M$:
\ba
R_{\sigma}=\frac{\sigma^{\uparrow\uparrow}}{\sigma^{\downarrow\uparrow}}
=\frac{\varepsilon}{\tau_p}\,\frac{G_E^2}{G_M^2}=
\frac{\varepsilon}{\tau_p}\frac {R^2}{ \mu_p^2}.
\label{rat2}
\ea
The expression on the right-hand side of Eq. (\ref{rat2}) for $R_{\sigma}$
is quite frequently used in publications. For example, the authors of \cite{Qattan2015}
reported two formulas for the reduced cross sections for the $ep \to ep$ process  that
include $R_{\sigma}$; however, they seem to be unaware of its physical meaning.

Using the standard notation and replacing $\delta_f$ with $P_r$ and $\delta_1$ with $P_t$,
we rewrite Eq. (\ref{delta_f}) for the degree of longitudinal polarization of the final
proton in the alternative form
\ba
P_{r}= P_t (R_{\sigma}-1)/(R_{\sigma}+1).
\label{Asim_pt}
\ea
Inverting the connection in (\ref{Asim_pt}), we obtain an expression for $R^2$ as a
function of $P_r/P_t$:
\ba
R^2=\mu_p^2\, \frac{\tau_p}{\varepsilon}\,\frac{1+R_p}{1-R_p}, \; \;
R_p=\frac{P_r}{P_t}\,,
\label{Rp}
\ea
which can be used to extract the ratio $R$ in the proposed method
of the polarization transfer from the initial to the final proton.

We calculate the $Q^2$ dependence of the polarization
asymmetry $A$ (\ref{Asim}), the ratio of the cross sections $R_{\sigma}$
(\ref{rat2}), and the polarization transfer to the proton $P_r$ (\ref{Asim_pt})
for both cases where the dipole dependence is valid
($R=R_d$) and is violated ($R=R_j$):
\ba
\label{Rd}
&&~~~~~~~~~~~~~~~~~~~~~~~~~~R_d=1 ,    \\
&&R_j = \frac{1}{1+0.1430Q^2-0.0086Q^4+0.0072Q^6}.~~~~
\label{Rdj}
\ea
The formula for $R_{j}$ is taken from \cite{Qattan2015}; Kelly’s parametrization
\cite{Kelly2004} can be used instead.

\begin{figure*}[h!tpb]
\centering
\includegraphics[width=0.45\textwidth]{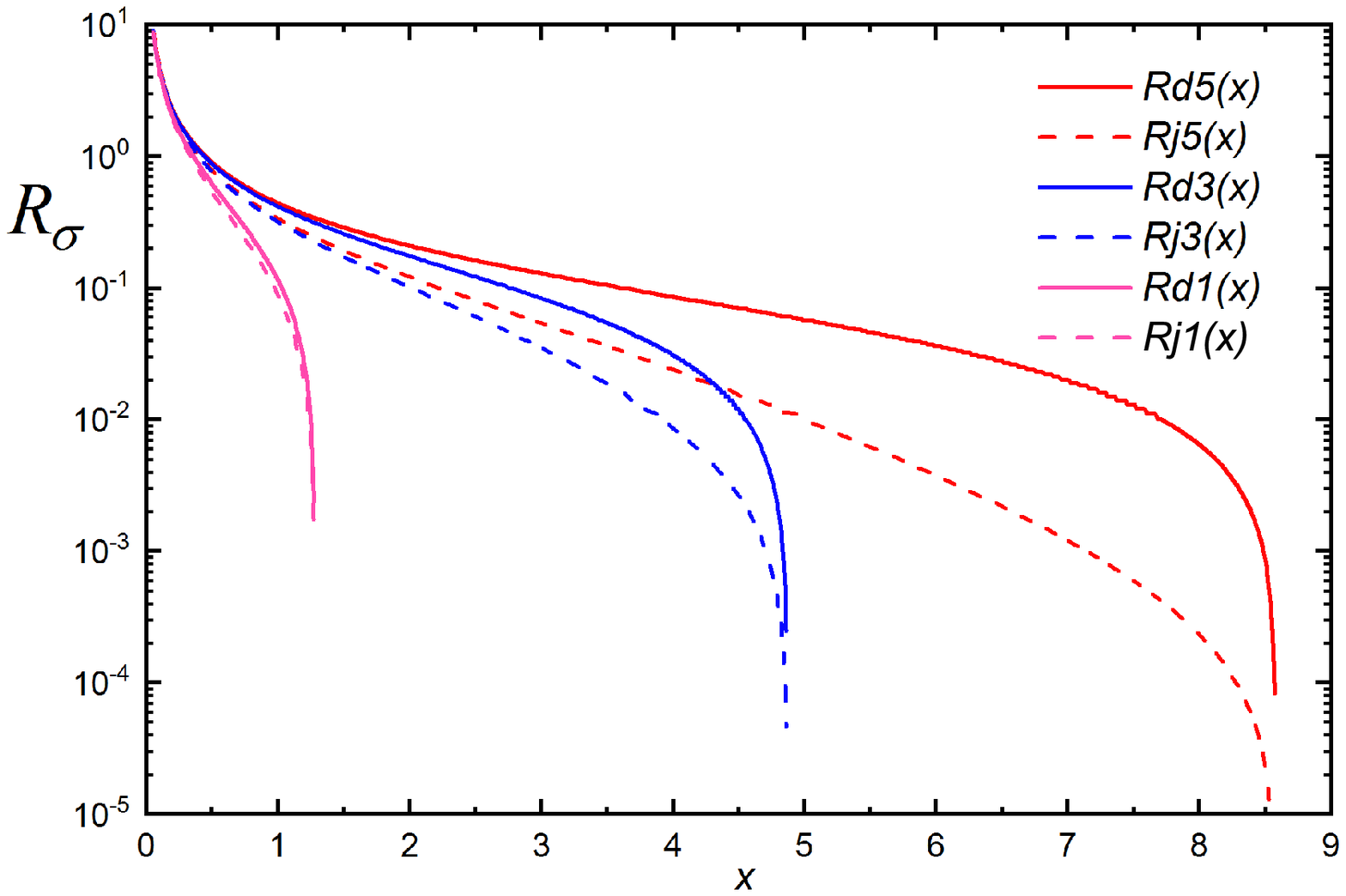}
\includegraphics[width=0.45\textwidth]{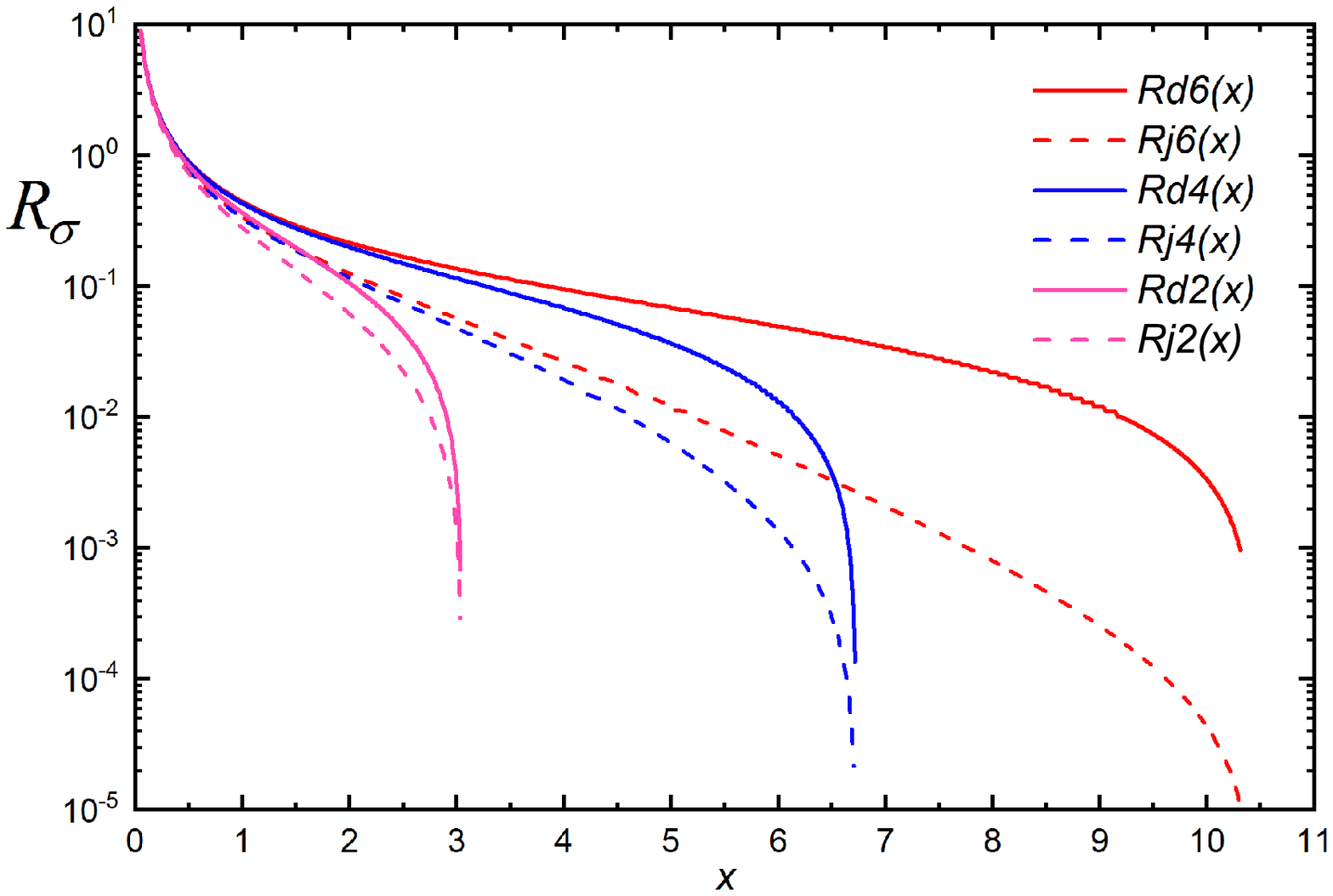}
\caption{
Ratio $R_{\sigma}$ (\ref{rat2}) calculated with
(lines $Rd1$, $ Rd2$,...,$ Rd6$) $R=R_d$ given by Eq. (\ref{Rd})
and (lines $Rj1$, $Rj2$, ...,$Rj6$) $R=R_j$ given by Eq. (\ref{Rdj})
versus $Q^2$ (GeV$^2$) for the energies $E_1=1, 2,...,6$ GeV, respectively.
}
\label{Ratfig}
\end{figure*}

\begin{figure*}[h!tpb]
\centering
\includegraphics[width=0.44\textwidth]{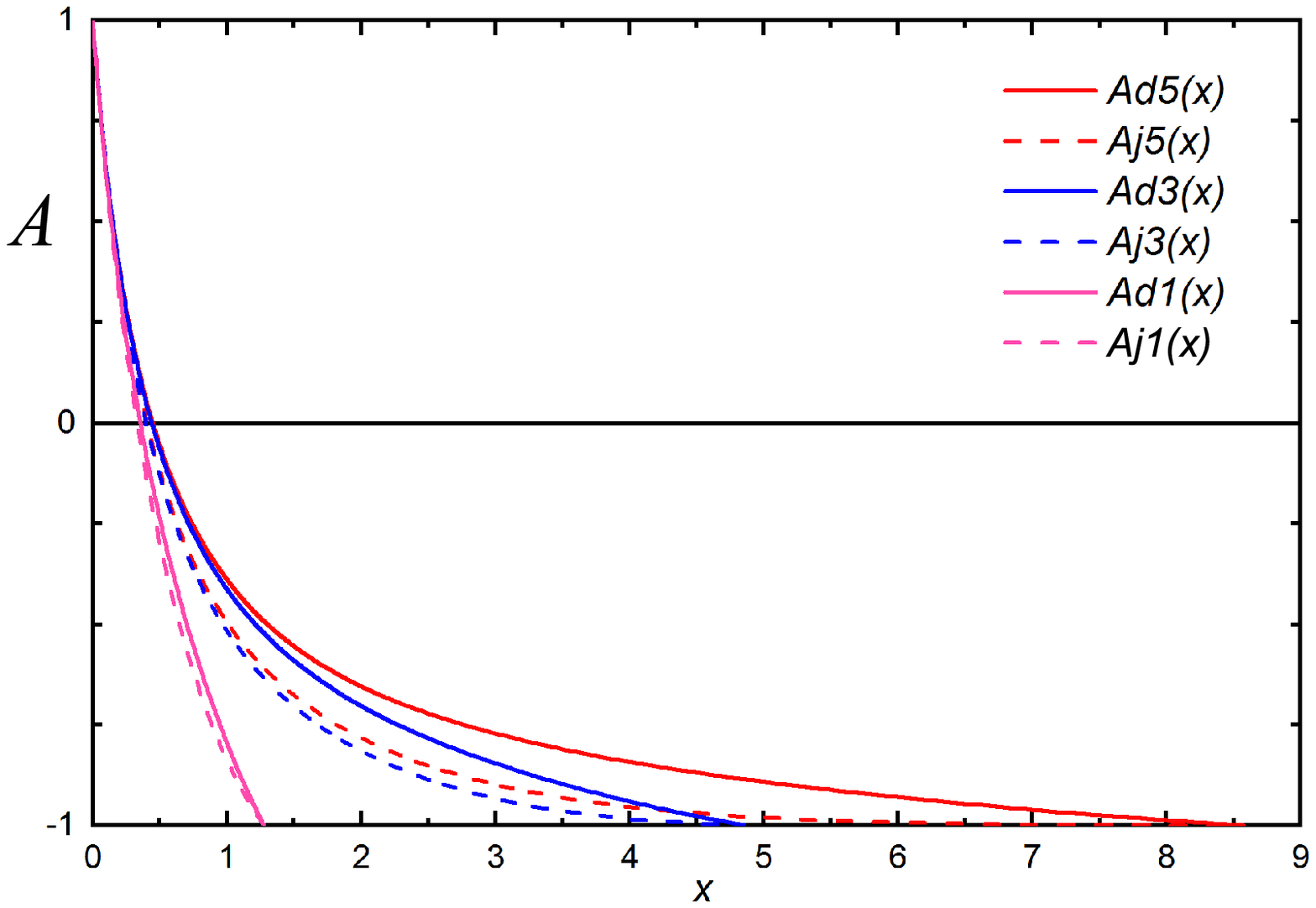}
\includegraphics[width=0.44\textwidth]{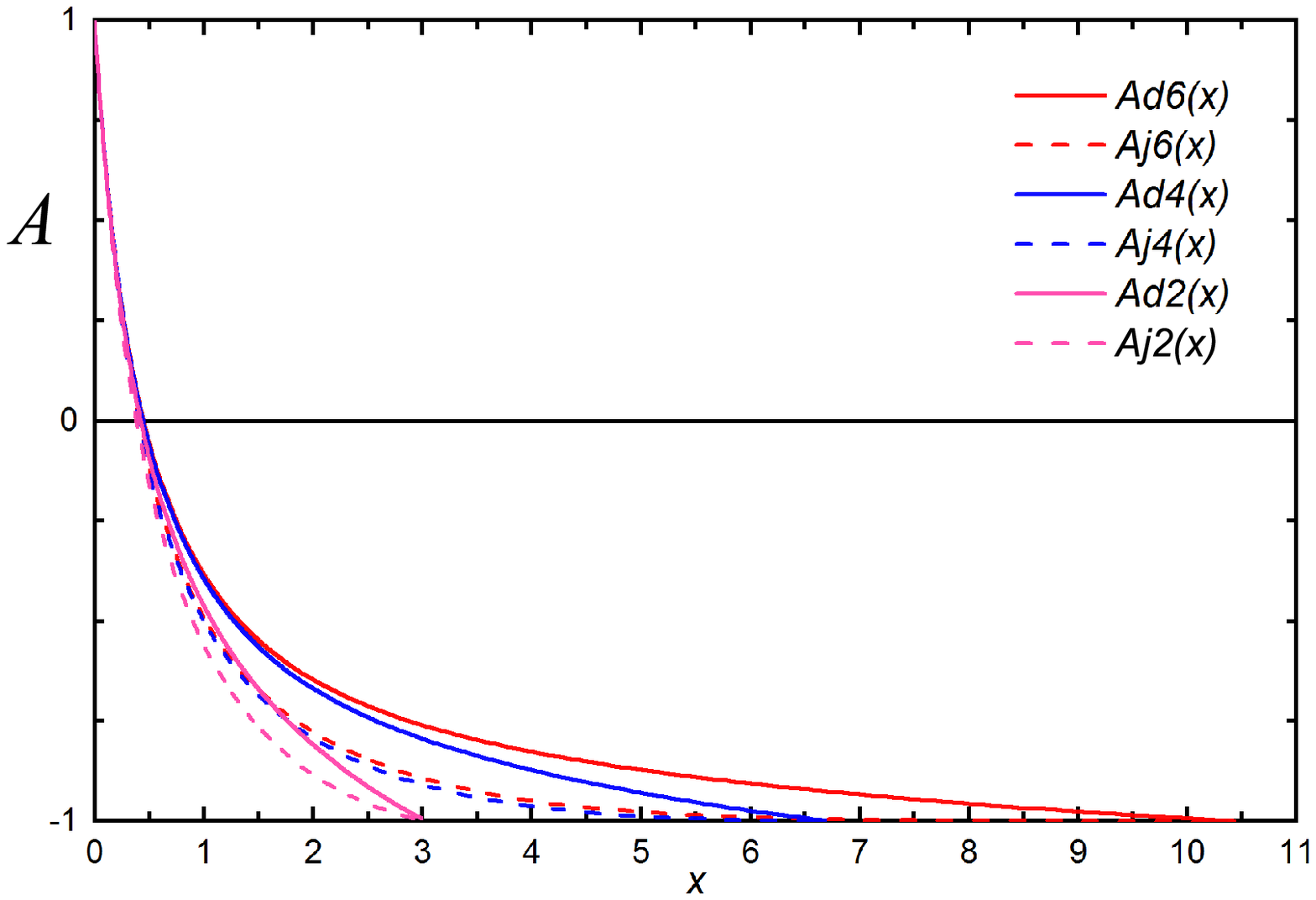}
\caption{
Polarization asymmetry $A$ (\ref{Asim}) calculated
with (lines $Ad1$, $Ad2$,...,$Ad6$) $R=R_d$  given by Eq. (\ref{Rd})
and (lines $Aj1$, $Aj2$,...,$Aj6$) $R=R_j$ given by Eq. (\ref{Rdj})
versus $Q^2$ (GeV$^2$) for the energies $E_1=1, 2, ...,6$ GeV, respectively.
}
\label{Asimfig12}
\end{figure*}

{\bf Numerical results and discussion.}
To identify general characteristics, Figs. \ref{Ratfig} and \ref{Asimfig12} show the $x = Q^2$
dependencies of the ratio of cross sections $R_{\sigma}$ (\ref{rat2}) and polarization asymmetry $A$ (\ref{Asim}),
respectively, calculated numerically with (lines $Rd1, Rd2, \ldots, Rd6$ and $Ad1, Ad2, \ldots, Ad6)$
$R = R_d$ given by Eq. (\ref{Rd}) and (lines $Rj1, Rj2, \ldots, Rj6$ and $Aj1, Aj2, \ldots, Aj6$)
$R = R_j$ given by Eq. (\ref{Rdj}) for the electron-beam energies $E_1 = 1, 2, \ldots, 6$ GeV,
respectively.

The plots displayed in Fig. \ref{Ratfig} show that the ratio of
the cross sections with and without proton spin flip
$R_{\sigma}$ (\ref{rat2}) for all electron beam energies decreases with
increasing $Q^2$. However, this decrease at $R=R_j$ is
faster than in the case of dipole dependence ($R=R_d$)
owing to the presence of the denominator in Eq. (\ref{Rdj})
for $R_J$. It should also be noted that the difference in
the behavior of the ratio $R_{\sigma}$ (\ref{rat2}) calculated with $R=R_d$
and $R_j$ is insignificant at a low electron-beam energy.
%

As seen in Fig. \ref{Ratfig}, the $Q^2$ dependence of $R_{\sigma}$ for each
electron-beam energy has a sharp boundary at $Q^2_{max}$,
which is the maximum possible value of $Q^2$ that corresponds
to the backscattering of the electron, i.e., scattering
at $180^{\circ}$. The values $Q^2_{max}$ for each electron-beam
energy $E_1=1, 2, ..., 6$ GeV are presented in Table \ref{Table1},
which shows that $Q^2_{max}$ does not exceed 10.45 GeV$^2$ for
all energies under consideration.

\begin{table}[h!]
\caption{
Values of $Q^2_{max}$ that set the boundaries of the $Q^2$
spectra of $R_{\sigma}$ and the values of $(Q_0)_{\{d,j\}}^2$
at which $\sigma^{\uparrow\uparrow}=\sigma^{\downarrow\uparrow}$;
the polarization asymmetry (\ref{Asim}) is zero in this case
}
\label{Table1}
\tabcolsep=1.25mm
\hfil
\footnotesize
\centering
\begin{tabular}{|  l | c | c | c |  c |  c |  c | c |}
\hline
$ E_1$  (\rm{GeV}) & 1.0 & 2.0 &   3.0 &  4.0 & 5.0 & 6.0  \\
\hline
$ Q^2_{max}$ (GeV$^2$) & 1.277 & 3.040 & 4.868 & 6.718 & 8.578 & 10.443  \\
\hline
$ (Q_0^2)_d$ (GeV$^2$) & 0.358 & 0.424 & 0.435 & 0.446  & 0.446  & 0.446 \\
\hline
$ (Q_0^2)_j$ (GeV$^2$) & 0.336 & 0.380 & 0.391 & 0.402 & 0.402 & 0.402  \\
\hline
\end{tabular}
\end{table}

Table \ref{Table1} also displays the values of $(Q_0^2)_{\{d,j\}}$ that correspond
to the coinciding cross sections with and without proton spin flip; in this case,
their ratio is $R_{\sigma}=1$, and the polarization asymmetry is zero. In the
case of dipole dependence, $(Q_0^2)_d \approx M^2/2$, where $M$ is
the mass of the proton. If the dipole dependence is
violated, $(Q_0^2)_j \approx 0.40$ GeV$^2$; i.e., the cross sections
$\sigma^{\uparrow\uparrow}$ and $\sigma^{\downarrow\uparrow}$ become
equal at approximately the same point where the linear decrease in the ratio $R$ begins.
Thus, the points where $Q^2=Q_0^2$ are in a certain sense specific.

If $Q^2> Q_0^2$, the spin-flip cross section $\sigma^{\downarrow\uparrow}$ exceeds
the cross section without spin flip $\sigma^{\uparrow\uparrow}$, and their ratio
is then $R_{\sigma}<1$. As a result, the helicity carried away by
the recoil proton becomes negative. Its absolute value reaches a maximum
value of 1 for the backscattered electron.

The calculations represented in Fig. \ref{Ratfig} make it possible
to understand why measurements of the ratio $R$
using the RT at high $Q^2$ values are difficult. They
should be conducted in a kinematics where the relative
contribution of the $\sigma^{\uparrow\uparrow}$ term to the cross section
$\sigma_R$ (\ref{Ross}) exceeds the accuracy $\Delta_0$ with which the Rosenbluth
cross section is measured in this experiment:
\ba
\frac{\sigma^{\uparrow\uparrow}}
{\sigma^{\uparrow\uparrow}+ \sigma^{\downarrow\uparrow}}
=\frac {R_{\sigma}}{(1+R_{\sigma})} > \Delta_0.
\label{treb1}
\ea
The restrictions on the kinematics of experiments where the RT is used have not yet been
considered in publications, including \cite{Egle_2018,Gramolin2016,Blunden2020}.
This problem is nevertheless of importance and deserves special attention.

According to Eq. (\ref{treb1}), the kinematics of the conducted
experiment should satisfy the condition
\ba
R_{\sigma} > \Delta_0,
\label{treb3}
\ea
which can be considered as a necessary condition for
reliable measurements and can be used as a criterion
for the reliability of measurements.

\begin{table*} [t!]
\small
\caption{
Values of $R_{\sigma}$ (\ref{rat2}) calculated with $R=R_d$ given by Eq. (\ref{Rd})
for the electron-beam energies $E_1=1, 2, ...,6$ GeV
and $Q^2=1, 2, ...,9$ GeV$^2$
}
\label{Table2}
\tabcolsep=1.25mm
\centering
\begin{tabular}
{| c| c|  c| c|  c|  c|  c|  c|  c|  c|}
 \hline
$Q^2$ (GeV$^2$) & 1.0 & 2.0 & 3.0 & 4.0 & 5.0 & 6.0 & 7.0 & 8.0 & 9.0  \\
\hline
$Rd6$, 6 GeV & 0.444 & 0.215 & 0.136 & 0.095 &  0.068 & 0.049 &  0.034 &  0.022 &  {0.012} \\
\hline
$Rd5$, 5 GeV & 0.440 & 0.209 & 0.129 & 0.086 & 0.057 & {0.036} &  0.020 & 0.006 & \\
\hline
$Rd4$, 4 GeV & 0.432 & 0.199 & 0.115 & 0.068 &  {0.037} &  {0.013} & & &  \\
\hline
$Rd3$, 3 GeV & 0.415  & 0.175 & 0.084 &  {0.031} &  & & & & \\
\hline
$Rd2$, 2 GeV & 0.365 & 0.105 & & & & & & & \\
\hline
$Rd1$, 1 GeV & 0.114 & & & & & & & & \\
\hline
\end{tabular}
\end{table*}

\begin{table*}[h!ptb]
\centering
\caption{
\label{Table3}
Values of $R_{\sigma}$ (\ref{rat2}) calculated with $R=R_d$ given by Eq. (\ref{Rd})
for $E_1$ and $Q^2$ values used in the measurements in \cite{Andivahis1994}}
\tabcolsep=1.25mm
\small
\begin{tabular}{|c| c| c| c|  c| c| c| c|  c|}
\hline
Q$^2$ (GeV$^2$)& 1.75 & 2.50 & 3.25 & 4.00 & 5.00 & 6.00 & 7.00 & 8.83 \\
\hline
$Rd9$, 9.800 GeV & & &  & 0.107 & 0.083 & 0.067 & 0.055 & \\
\hline
$Rd8$,  5.507 GeV & 0.246 & 0.165 & 0.120 & 0.091 & 0.064 & & & {\bf 0.006} \\
\hline
$Rd7$,  4.507 GeV & & &  & 0.079 & 0.049 & & {\bf 0.009} & \\
\hline
$Rd6$,  3.956 GeV & & 0.157 &  0.100 & 0.067 &  0.035 &  {\bf 0.012} & & \\
\hline
$Rd5$, 3.400 GeV & & 0.136 & 0.085 & 0.049 & {\bf 0.016} & & & \\
\hline
$Rd4$,  2.837 GeV & & 0.114 & 0.059 &  {0.022} & & & & \\
\hline
$Rd3$,  2.407 GeV & 0.182 & 0.087 &  {0.029} & & & & & \\
\hline
$Rd2$,  1.968 GeV & & 0.041 & & & & & & \\
\hline
$Rd1$,  1.511 GeV & 0.065 & & & & & & & \\
\hline
\end{tabular}
\end{table*}

Table \ref{Table2} summarizes the values of $R_{\sigma}$ (\ref{rat2}) calculated
with $R=R_d$ for $E_1=1,2,\ldots, 6$ GeV and $Q^2=1,2,\ldots, 9$ GeV$^2$ corresponding
to the $Rd1, Rd2,..., Rd6$ plots in Fig. \ref{Ratfig}. All $R_{\sigma}$ values
in Table \ref{Table2} for $Q^2=7.0$ and 8.0 GeV$^2$, except one where $R_{\sigma}=0.006$,
satisfy the inequality $R_{\sigma} \geqslant 0.020$. Using criterion (\ref{treb3}),
we conclude that the RT-based measurements at $Q^2=7.0$ GeV$^2$ should be carried out
with an accuracy of no less than 1.9 \%, while the measurements made at $Q^2=8.0$ GeV$^2$
require an accuracy of $0.3 \div 0.5$ \%. Thus, RT measurements of the ratio $R$
at high $Q^2$ values is difficult because the relative contribution of the
$\sigma^{\uparrow\uparrow}$ term to the Rosenbluth cross section (\ref{Ross})
decreases, and the accuracy of measuring this quantity should be correspondingly
increased. It is noteworthy that measuring the Rosenbluth cross sections with an
accuracy better than 2 \% using the RT technique was an unrealistic task in old experiments
for many reasons \cite{Bernauer2014}.

The criterion (\ref{treb3}) for the reliability of data does not specify the meaning of
the accuracy $\Delta_0$ of the measurement of the Rosenbluth cross sections and specific
inaccuracies (statistical, systematic, or normalization) determining this accuracy.
Analyzing the reliability of the experiment reported \cite{Andivahis1994} using results obtained in
\cite{Blunden2020}, we show below that $\Delta_0$ is determined by the normalization
uncertainty. After that, the same approach will be applied to analyze the measurements
reported in \cite{Qattan2005}.

{\bf Analysis of the reliability of two well-known experiments.}
To analyze the reliability of the ratio $R$ measured in the experiment in \cite{Andivahis1994},
we calculate the ratio $R_{\sigma}$ (\ref{rat2}) for all electron beam energies $E_1$
and squares of the momentum transfer to the proton $Q^2$ used in this experiment.
The corresponding results are presented in Table \ref{Table3}. Empty cells in Table \ref{Table3}
indicate that no measurements were made at the corresponding $E_1$ and $Q^2$ values.

The values displayed in bold in Table \ref{Table3} for $Q^2\geqslant 5.0$ GeV$^2$ are classified
as unreliable results. To be confident in this conclusion, we address study \cite{Blunden2020} where
experiments \cite{Qattan2005,Andivahis1994} were reanalyzed taking into account the contribution
from the TPE. Figure 15b in \cite{Blunden2020} shows that measurements at $Q^2 < 5.0$ GeV$^2$ in
\cite{Andivahis1994} with the added TPE contribution agree well with results from \cite{Puckett17};
however, even the inclusion of the  TPE fails to remove disagreements at $Q^2 = 5.0$ GeV$^2$.
For this reason, the lower value in the column for $Q^2 = 5.0$ GeV$^2$ in Table \ref{Table3}
is classified as unreliable, i.e., having insufficient accuracy.
Table \ref{Table3} and criterion (\ref{treb3}) show that the accuracy of measurements in \cite{Andivahis1994}
was about $1.7 \div 2.1$ \%. The normalization uncertainty in measurements of the Rosenbluth cross
section, which at all $Q^2$ values in \cite{Andivahis1994} was 1.77 \%
(see \cite{Andivahis1994,Gramolin2016,Bernauer2014}), falls into the same range. Thus, the accuracy
of measurements $\Delta_0$ appearing in Eq. (\ref{treb3}) should be identified with the normalization
uncertainty. Given this accuracy (1.77 \%), reliability criterion (\ref{treb3}) fails for all values
on the diagonal of Table \ref{Table3} at $Q^2\geqslant 5.0$ GeV$^2$.

The value in Table \ref{Table3} for $Q^2=8.83$ GeV$^2$ and $E_1=5.507$ GeV corresponds
to $R_{\sigma}=0.006$, a value that requires the accuracy of measurements of $0.3 \div 0.5$ \%.
However, this accuracy was attained in the experiment reported in \cite{Bernauer2010-1}
and performed only in 2010 and, moreover, in the region  $Q^2<1$ GeV$^2$. It should be noted that
the statistical, systematic, and normalization uncertainties of measurements at
$Q^2=8.83$ GeV$^2$ in \cite{Andivahis1994} are 3.89, 1.12, and 1.77 \%, respectively
\cite{Gramolin2016}. They significantly exceed the measurement accuracy required
at $Q^2=8.83$ GeV$^2$. In addition, the procedure of RT based
measurements was violated at $Q^2=8.83$ GeV$^2$,
since measurements should be performed for each $Q^2$
value for at least two or better three energies of the
electron beam \cite{Bernauer2010}.

The results of the reanalysis of the experiment \cite{Qattan2005} with the added TPE contribution
were also shown in Fig.15 b in \cite{Blunden2020}. They are systematically located above
the green stripe in that figure that corresponds to the results of polarization experiments
in \cite{Puckett17}, which at first glance is an indication that measurements \cite{Qattan2005}
are unreliable. However, this is not true. To analyze the reliability of the RT-based measurements
of the $R$ ratio in the experiment reported in \cite{Qattan2005}, the ratio $R_{\sigma}$ (\ref{rat2})
was calculated for all energies of the electron beam $E_1$ and squares of the momentum transfer
to the proton $Q^2$ used in the experiment. The corresponding results are presented in Table \ref{Qattan}.

\begin{table}[h!]
\caption{
Values of $R_{\sigma}$ (\ref{rat2}) calculated with $R=R_d$ given by
Eq. (\ref{Rd}) for $E_1$ and $Q^2$ values used in the measurements in
\cite{Qattan2005}
}
\label{Qattan}
\small
\centering
\tabcolsep=1.25mm
\begin{tabular}
{| l |  c | c |  c | }
\hline
$Q^2$ (\rm{GeV}$^2$)  & 2.64 & 3.20 & 4.10  \\
\hline
$Rd5$, $E_1=4.702$ GeV & 0.148 & 0.115 & 0.078 \\
\hline
$Rd4$, $E_1=3.772$ GeV & 0.134  & 0.098 & 0.058  \\
\hline
$Rd3$, $E_1=2.842$ GeV & 0.102 & 0.063 & {\bf 0.018}  \\
\hline
$Rd2$, $E_1=2.262$ GeV  & 0.061 & {\bf 0.018}  &     \\
\hline
$Rd1$, $E_1=1.912$ GeV  &  {\bf 0.020} &  &       \\
\hline
\end{tabular}
\end{table}

The accuracy of measurements required for the values given in bold in Table \ref{Qattan} proves
to correspond to the normalization accuracy in \cite{Qattan2005} equal to 1.7 \% \cite{Bernauer2014},
thus indicating their reliability. This conclusion is based on the necessity of using
the same approach to analyze measurements in \cite{Qattan2005} and \cite{Andivahis1994},
where $\Delta_0$ is determined by the normalization uncertainty. Since measurements
in \cite{Qattan2005} are reliable, a more accurate reanalysis is required to remove
the remaining disagreements between “\cite{Qattan2005} + TPE” and \cite{Puckett17}
discovered in \cite{Blunden2020}.

{\bf On the feasibility of the measurement of the ratio of SFFs in the $e \vec p \to e \vec p$ process.}
The method to measure squares of SFFs in the processes with and without proton spin flip proposed in
\cite{JETPL18} requires a fully polarized proton target, which may be obtained in the distant future.
As stated above, it may be considered in a broader sense as a technique based on the polarization
transfer from the initial to the final proton.
Generally, if the initial proton is partially polarized, the degree of the longitudinal
polarization transfer to the recoil proton is given by Eq. (\ref{Asim_pt}). An experiment
to measure this quantity is currently quite realistic, since a partially polarized
proton target with a high degree of polarization $P_t=(70 \pm 5)$ \%  was already used
in \cite{Liyanage2020}. For this reason, it would be the most expedient
to carry out the proposed experiment at the facility used by the SANE collaboration
\cite{Liyanage2020} at the same value of $P_t=0.70$, electron beam energies $E_1=4.725$ and
5.895 GeV, and the same values of the square of the momentum transfer to the proton
$Q^2 = 2.06$ and 5.66 GeV$^2$. The difference between the proposed experiment and that
performed in \cite{Liyanage2020} is that the electron beam should be unpolarized,
while the detected recoil proton should move exactly along the direction of quantization
of the proton target spin. The degrees of the longitudinal and transverse polarizations of the
final proton were measured in \cite{Jones00,Gay01,Gay02,Pun05,Puckett10,Puckett12,Puckett17}.
In the proposed experiment, the degree of the longitudinal polarization of the recoil proton
alone should be measured, which is an advantage compared to the technique proposed in \cite{Rekalo74}.

\begin{figure}[h!]
\centering
\includegraphics[width=0.45\textwidth]{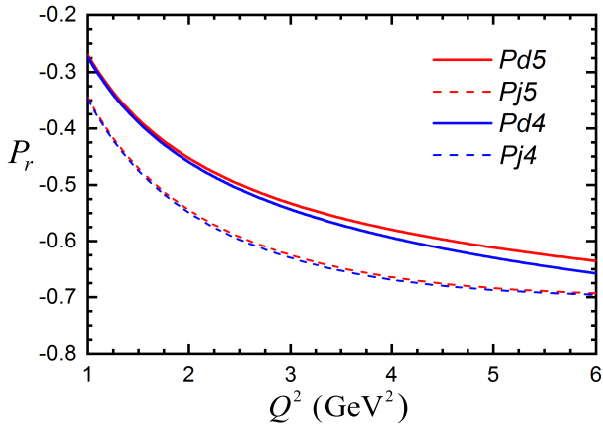}\\
\caption{
Longitudinal polarization of the recoil proton $P_r$ (\ref{Asim_pt}) calculated with
(lines $Pd5$ and $Pd4$) $R=R_d$ given by Eq. (\ref{Rd}) and (lines $Pj5$ and $Pj4$) $R=R_j$
given by Eq. (\ref{Rdj}) versus the square of the momentum transfer to the proton $Q^2$ (GeV$^2$)
for the values $E_1$ and $P_t$ used in \cite{Liyanage2020}.
}
\label{exp}
\end{figure}

The calculated $Q^2$ dependence of the polarization transfer to the final proton
$P_r$ (\ref{Asim_pt}) in the kinematics of the experiment reported \cite{Liyanage2020}
is plotted in Fig. \ref{exp}, where the lines $Pd5$ ($Pd4$) and $Pj5$ ($Pj4$) are obtained
with $R=R_d$ given by Eq. (\ref{Rd}) and with $R=R_j$  given by Eq. (\ref{Rdj})
for the electron beam energy $E_1=5.895$ (4.725) GeV, respectively. For all lines
in Fig. \ref{exp}, the degree of polarization of the proton target is $P_t=0.70$.
%

The plots in Fig. \ref{exp} show that the polarization transfer to the final proton significantly
depends on the form of the $Q^2$ dependence of $R$. The absolute value of this polarization
transfer in the case of the violation of the SFFs scaling, i.e., in the case $R=R_j$, is significantly
greater than that in the case $R=R_d$. A quantitative estimate of this difference is shown
in Table \ref{Table5}, where the longitudinal polarizations of the recoil proton $Pj5$, $Pd5$, $Pj4$,
and $Pd4$, as well as their relative differences $\Delta_{dj5}=(Pj5-Pd5)/Pj5$
and $\Delta_{dj4}=(Pj4 - Pd4)/Pj4$ in percent, are displayed for two electron-beam energies
of 5.895 and 4.725 GeV and two values $Q^2= 2.06$ and 5.66 GeV$^2$.

\begin{table}[h!]
\centering
\caption{
Longitudinal polarization of the recoil proton $P_r$ (\ref{Asim_pt}) for electron-beam
energies of 5.895 and 4.725 GeV and $Q^2 = 2.06$ and 5.66 GeV$^2$
}
\label{Table5}
\tabcolsep=1.25mm
\small
\begin{tabular}
{| c | c | c | c | c | c | c | c |}
\hline
$Q^2$ (\rm{GeV}$^2$) & $Pd5$ & $Pj5$ & $Pd4$ & $Pj4$ & $\Delta_{dj5}$, \%  & $\Delta_{dj4}$, \%  \\
\hline
2.06 & --0.46 & --0.55 & --0.47 & --0.56 & 16.6 & 16.1 \\
\hline
5.66 & --0.63 & --0.69 & --0.65 & --0.69 & 9.1 & 6.4   \\
\hline
\end{tabular}
\end{table}

Table \ref{Table5} shows that the relative differences between $Pj5$ and $Pd5$
and between $Pj4$ and $Pd4$ at $Q^2=2.06$ GeV$^2$ are 16.6 and 16.1 \%, respectively,
and decrease to 9.1 and 6.4 \%, respectively, at $Q^2 = 5.66$ GeV$^2$.
%

{\bf Conclusions.}
Using the results of the JLab’s polarization experiments, where the ratio $R$  was measured
in the $\vec e  p \to e \vec p$ process, we have numerically analyzed the ratio of
cross sections with and without proton spin flip and the polarization asymmetry
as functions of the square of the momentum transfer to the proton in this process
for the case where the initial (at rest) proton and the final proton are fully polarized
and have a common spin quantization axis that coincides with the direction
of motion of the detected recoil proton. If the initial proton is partially polarized,
the longitudinal polarization transfer to the proton is calculated for the kinematics
used by the SANE collaboration \cite{Liyanage2020} in the experiments to measure double
spin asymmetry in the $\vec e \vec p \to e p$ process. A noticeable sensitivity of the polarization
transfer to the proton to the form of the $Q^2$ dependence of the ratio $R$ has been found.
This sensitivity may be used to conduct a new independent experiment to measure this dependence
in the $ e \vec p \to e \vec p$ process. A criterion for the reliability of measurements
of $R$ using the Rosenbluth technique has been proposed and used to analyze results of two
experiments reported in \cite{Andivahis1994,Qattan2005}. The results of the analysis may
be used to identify the reasons for the disagreements that remain between the results
of measurements \cite{Qattan2005} with the added TPE contribution and polarization experiments
\cite{Puckett17}, which were discovered in \cite{Blunden2020}.

{\bf Acknowledgments.}
I am grateful to R. Lednicky for his interest in the study and helpful discussions.

\end{document}